\pdfoutput=1
\documentclass[aps,twocolumn,superscriptaddress, prl]{revtex4-2}
\usepackage{graphicx}
\usepackage{dcolumn}
\usepackage{bm}
\usepackage{amsmath}
\usepackage{amssymb}
\usepackage{braket}
\usepackage[caption=false]{subfig}
\usepackage[colorlinks=true]{hyperref}
\usepackage{microtype}
\usepackage{comment}
\usepackage{xcolor}
\usepackage{physics}
\usepackage{mathrsfs}
\usepackage{comment}

\usepackage{soul}

\def\*#1{\mathbf{#1}}
\def\mc#1{\mathcal{#1}}
\def\mb#1{\mathbb{#1}}

\def\t#1{\text{#1}}

\def\bs#1{\boldsymbol{#1}}
\def\tt#1{\textit{#1.}|}

\def\grad{\bs\nabla}

\begin{document}
\title{Nonabelian vortices: From topological strings to nonvolatile memory}

\author{Chau Dao}
\thanks{These authors contributed equally to this work.}
\affiliation{Department of Physics and Astronomy and Bhaumik Institute for Theoretical Physics, University of California, Los Angeles, California 90095, USA}

\author{Eric Kleinherbers}
\thanks{These authors contributed equally to this work.}
\affiliation{Department of Physics and Astronomy and Bhaumik Institute for Theoretical Physics, University of California, Los Angeles, California 90095, USA}

\author{Yaroslav Tserkovnyak}
\affiliation{Department of Physics and Astronomy and Bhaumik Institute for Theoretical Physics, University of California, Los Angeles, California 90095, USA}


\date{\today}

\begin{abstract}
We investigate the topological classification and properties of nonabelian vortices that arise in a 2-dimensional film of a two-component condensate. For a repulsive interaction between the condensates, we realize an order parameter in the ordered configuration space of three complex numbers $\mc C_3(\mb C)$. Analyzing the homotopic properties, we find the emergence of elementary nonabelian vortex-antivortex pairs connected by energetic strings that are rooted in the repulsive interaction and carry the nonabelian information. Moreover, we predict the existence of Brunnian vortices, which are textures exhibiting zero net winding of the condensate phases and can only be detected by a higher-order winding number. Finally, to motivate technological applications rooted in nonabelian vortex transport, we devise a nonvolatile memory storage device that encodes topologically robust information.
\end{abstract}

\maketitle

\tt{Introduction}Numerous topological defects such as vortices, skyrmions, hopfions, and hedgehogs are rooted in Abelian homotopy groups and have garnered significant interest in condensed matter physics. For such defects, the notions of transport and extensive build-up of topological charge have a natural interpretation since the charge is $\mb Z$-valued. However, for certain order parameter spaces, a 2-dimensional system can host \textit{nonabelian} vortices. Compared to their abelian counterparts, nonabelian vortices have an enriched internal structure, giving rise to orientation-dependent combination rules~\cite{merminRMP79,wuPRX25} as well as being endowed with a higher capacity to encode information. However, they can no longer be faithfully classified by an integer-valued charge, so the conventional topological conservation laws underpinning the hydrodynamics of $\mb Z$-valued topological charge~\cite{tserkovnyakJAP18,zouPRB19,zouPRL20,daoPRB25,daoPRB26,rybakovPRB25, rybakovARX22} lose their meaning. This presents a challenge for understanding how nonabelian vortices can be classified, nucleated, and transported. 

In this Letter, we address these problems by investigating an illustrative case study of nonabelian vortices. We consider a 2-dimensional system in which at every point, the order parameter can be described by a tuple of three ordered complex numbers $\bs{\Psi}(\vb{r})=\left(\psi_0(\*r), \psi_1(\*r),\psi_2(\*r)\right)$ with $\psi_i(\*r)\neq \psi_j(\*r)$ for $i\neq j$ and $i,j\in\{0,1,2\}$, forming the ordered configuration space of three distinct complex numbers $ \mathcal C_3(\mb C)$. 
Along a loop in real space enclosing a vortex, the complex numbers can braid around each other, which is captured by the fundamental homotopy group $\pi_1\left[ C_3(\mb C)\right]=P_3$. Here, $P_3$ is the nonabelian pure braid group and defects contained within a closed loop are, in turn, classified by the conjugacy classes of $P_3$~\footnote{In contrast, defects stemming from an order parameter consisting of only \textit{two} ordered complex numbers are classified by the winding number, $\pi_1\left[ C_2(\mb C)\right]=\mb Z$, which simply counts the winding of $\psi_1$ around $\psi_0$ along a closed loop. }.

\begin{figure}
    \centering 
    \includegraphics[width=\linewidth]{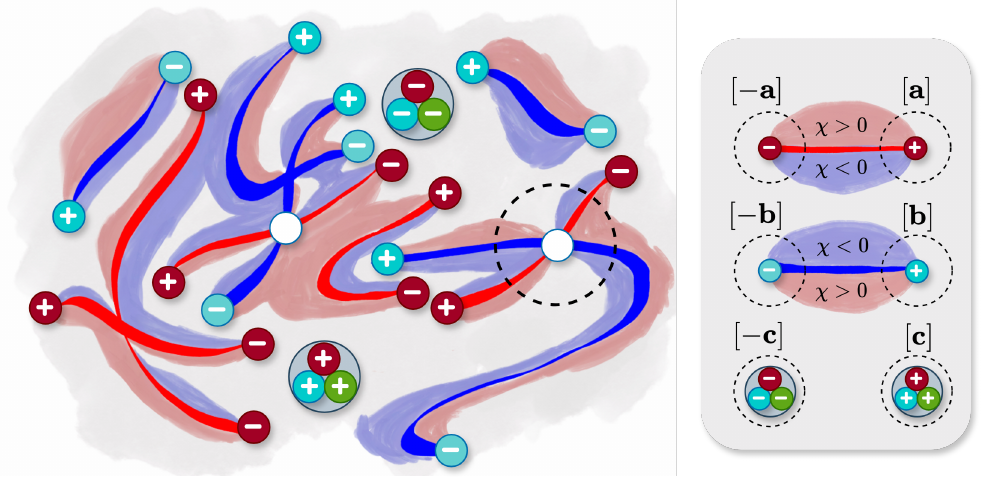}
    \caption{Illustration of the soup of vortex-antivortex pairs with a legend labeling the elementary vortex-antivortex pairs on top of the ground state (gray). The red, green, and blue  circles inscribed with plus and minus signs denote singular vortex cores with relative winding numbers $n_{01}=\pm1$, $n_{12}=\pm1$, and $n_{20}=\pm1$, respectively. The light red (blue) shaded regions indicate a chirality $\chi=+1$ ($\chi=-1$), where the complex numbers $\psi_0$, $\psi_1$, and $\psi_2$ are ordered  counterclockwise (clockwise) in the complex plane. Solid red and blue strings connecting the vortex cores correspond to excited collinear configurations with chirality $\chi=0$, where $\psi_1$ and $\psi_2$ are located in the center, respectively. At the intersection of blue and red strings, indicated by the white circles, there must be a singularity with zero winding number, which is emblematic of the Brunnian vortex enclosed by the black dashed circle.}
    \label{fig:1}
\end{figure}

\begin{figure*}
    \centering    
    \includegraphics[width=0.9\linewidth]{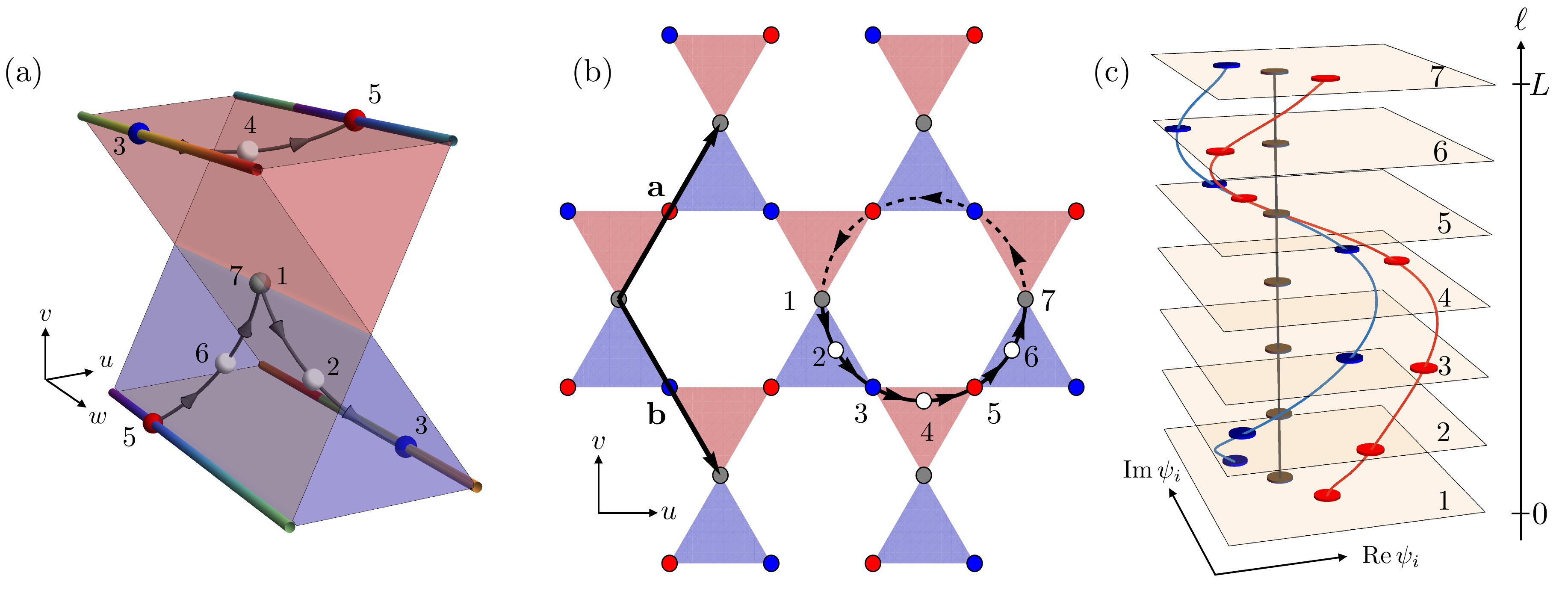}
    \caption{(a) Unit cell of order parameter space, (b) extended order parameter space projected to the  plane orthogonal to the $(1,1,1)$ direction, and (c) texture of $\psi_0$ (gray), $\psi_1$ (red) and $\psi_2$ (blue) on a loop $\partial \cal A$ with $0 \le \ell<L$ encircling a $[\vb{b},\vb{a}]$ vortex. In (a)-(b), we plot along the high symmetry directions by introducing the coordinates $u=\theta_{01}-2\theta_{12}+\theta_{20}$, $v={\sqrt{3}}(\theta_{01}-\theta_{20})$, and $w=\theta_{01}+\theta_{12}+\theta_{20}$. Note that in $(u,v,w)$ coordinates the primitive vectors are $\vb{a}=2\pi(1,\sqrt{3},1)$, $\vb{b}=2\pi(1,-\sqrt{3},1)$, and $\vb{c}=6\pi(0,0,1)$, so they have an out-of-plane component along $w$. The numbers $n\in \{1,\ldots,7\}$ indicate distinct states $\bf{\Psi}$  parametrized via $(u,v,w)=(2\pi\cos (\pi+\alpha_n),2\pi\sin (\pi+\alpha_n),4\alpha_n)$ with $\alpha_n=\pi (n-1)/6$. }
    \label{fig:2}
\end{figure*}

As a possible realization of such an order parameter $\bs{\Psi}$, we propose to use  two-component condensates described by the complex fields $\psi_1$ and $\psi_2$ which have been subject to intensive theoretical and experimental studies. This includes exciton-polariton condensates with two chiral polarizations~\cite{liPRB15},  Bose-Einstein condensates in cold atomic systems using two hyperfine states~\cite{sonPRA02,kasamtsuPRL04}, magnon condensates at different wave numbers~\cite{dzyapkoPRB17}, and also two-component superconductors~\cite{leggettPTP66,babaevPRL02,babaevNAT09,tanakaPRL01}.
In particular, we require two ingredients. First, a Mexican hat potential $V(\psi_i)$ with $i\in\{1,2\}$ that prohibits each condensate field from vanishing. This allows us to define  $\psi_0 \equiv 0$ as an artificial complex field, with $\psi_1\neq \psi_0$ and $\psi_2\neq \psi_0$. Second, a repulsive interaction $U(\psi_1,\psi_2)$ between the condensates that enforces $\psi_1\neq \psi_2$. A similar interaction, albeit attractive, was studied in Ref.~\cite{sonPRA02}.

Choosing to remain agnostic of the specific physical realization of the two-component condensate, we expect on topological grounds that such a system, when excited (e.g. thermally) far away from the trivial ground state (given by ${\psi_1}=-{\psi_2}$ due to repulsion), will resemble a soup of vortex-antivortex pairs, see Fig.~\ref{fig:1}. The soup consists of three distinct species of elementary vortex-antivortex pairs, which are henceforth denoted by $[\*v]$ for vortices and $[{-}\*v]$ for antivortices, with $\*v\in\{\*a,\*b,\*c\}$ being the species. The $[\*c]$ vortices are abelian, while $[\*a]$ and $[\*b]$ are nonabelian.

The topological constraints set by the $\mc C_3(\mb C)$ order parameter space can be distilled to a set of three rules for the vortices shown in Fig.~\ref{fig:1}. First, the vortices have charges $n_{01}$, $n_{12}$, and $n_{20}$ (indicated by red, green, and blue circles, respectively) given by the relative winding number $n_{ij}$ of $\psi_i$ around $\psi_j$ for $i\neq j$ as we enclose a vortex [see Eqs.~\eqref{eq:relangles}-\eqref{eq:relwindingnumbers} below]. The nonabelian vortices $[\*a]$ and $[\*b]$ carry only the condensate winding numbers $n_{01}=+1$ and $n_{20}=+1$, respectively. On the other hand, the abelian vortex  $[\*c]$ carries all winding numbers $n_{01}=n_{12}=n_{20}=1$, as it corresponds to jointly winding $\psi_1$ and $\psi_2$ about $\psi_0$~\footnote{It is interesting to note that upon turning off the $U(\psi_1,\psi_2)$ interaction, the $[\*a]$ and $[\*b]$ vortices revert to the conventional abelian $\t U(1)$ vortices of the $\psi_1$ and $\psi_2$ condensates, respectively. Furthermore, $[\*c]$ reverts to its constituent condensate vortices, which are now decoupled, since the $n_{12}$ winding number is no longer protected.}. The respective antivortices carry opposite winding. Second, the vortex and antivortex of a given pair $\{[\vb{a}],[{-}\vb{a}]\}$ and $\{[\vb{b}],[{-}\vb{b}]\}$ are connected by red and blue ``topological strings," respectively, which indicate regions that deviate from  $\psi_1=-\psi_2$  and are penalized by repulsion $U(\psi_1,\psi_2)$. The latter are crucial since they encode the nonabelian information. Notably, the strings are nonsingular, while the only true singularities are at the vortex cores. Third, under continuous deformation, strings  of the same color can be crossed while strings of different color cannot. Crossing strings of different colors requires introducing a singularity with zero winding number, $n_{ij}=0$, at the intersection (indicated by white circles). 

In the following, we first propose a minimal model which realizes the $\mc C_3(\mb C)$ order parameter space and elucidate how the topological constraints detailed above come about. As an example of the striking topological features that may arise, we predict the existence of exotic topologically robust vortices that carry zero net winding (coined Brunnian vortices) and motivate a higher-order winding number that can detect them. Moreover, we conceptually motivate potential applications for this physics by devising a nonvolatile memory storage concept on which we can write and transmit information.

\tt{Minimal model}The two-dimensional film of a two-component condensate with $\psi_1(\vb{r})$ and $\psi_2(\vb{r})$ is phenomenologically described by the free energy 
\begin{align}\label{eq:free_energy}
\begin{split}
    \mathcal F[\psi_1,\psi_2] = \int \t dA\, \Bigg \{\sum_{i=1,2}&\left[ \frac{A}{2} \abs{\grad\psi_i}^2  + V(\psi_i) \right] \\& + U(\psi_1, \psi_2)\Bigg\},
\end{split}
\end{align}
where $A$ describes the condensate stiffness and $U(\psi_1, \psi_2)=-U_0 \abs{\psi_1-\psi_2 }^2$ for $U_0>0$ is the repulsive interaction between the two condensates to energetically enforce $\psi_1(\vb{r})\neq \psi_2(\vb{r})$. Moreover, the Mexican hat potential $V(\psi_i) = -\alpha|\psi_i|^2 + \beta|\psi_i|^4$ (where $\alpha,\beta>0$ are phenomenological parameters) prohibits the two condensate fields from vanishing. Hence, we introduce $\psi_0(\*r) \equiv 0$ such that $\psi_i(\*r)\neq \psi_j(\*r)$ for $i\neq j$, where $i\in\{0,1,2\}$~\footnote{To better protect the order parameter space, a more nonlinear or localized repulsion can be used. For example, a flat rim of the Mexican hat potential paired with a sufficiently local repulsion that diverges for $\psi_1=\psi_2$ would be ideal.}. As discussed above, this forms the order parameter $\bs{\Psi}=(\psi_0,\psi_1,\psi_2)\in\mathcal C_3(\mb C)$. 
The ground state fulfills $\psi_1=-\psi_2$ and can be written as $\bs{\Psi}_0=\sqrt{(U_0+\alpha/2)/\beta}(0,e^{i\varphi},-e^{i\varphi})$ with $\varphi$ being a spontaneously chosen phase. In the following, we discuss topological excitations of this ground state.

\tt{Order parameter space}To determine the geometry of the order parameter space, we can first reduce the number of degrees of freedom, retaining only those necessary to describe a topologically distinct braid. For this reason, we mod out translations and scaling of the complex fields $\psi_i$ and parametrize the remaining degrees of freedom by the three relative phases
\begin{align}
\theta_{ij}\equiv \arg(\psi_i-\psi_j)\in \mathbb{R}.\label{eq:relangles}
\end{align}
Along a closed loop $\partial \cal A$ parametrized by $\ell$, the winding of $\psi_i$ around $\psi_j$ is a topologically conserved quantity:
\begin{align}
    \int_{\partial \cal A} \mathrm{d} {\ell}\,  \partial_\ell{\theta}_{ij} = 2\pi n_{ij},\label{eq:relwindingnumbers}
\end{align} 
where $n_{ij}\in\mathbb{Z}$ are the integer-valued relative winding numbers.

The order parameter space of allowed configurations  $(\theta_{01},\theta_{12},\theta_{20})$ is shown in Fig.~\ref{fig:2}(a)-(b)~\footnote{The configuration space $\mathcal C_3(\mb C)$ can be understood as an effective order parameter space, in the sense that it parametrizes the low-energy manifold of the system. However, different configurations $\bs \Psi$ do not generally have the same energy.}. The space is  translationally invariant along the $(1,1,1)$ direction, while the plane orthogonal to $(1,1,1)$ is periodically perforated by  hexagonal holes, see the white regions in Fig.~\ref{fig:2}(b). To clarify the meaning of these forbidden regions, let us first visualize an arbitrary point in the configuration space $\mc C_3(\mb C)$ as a triangle in the complex plane spanned by $\psi_0$, $\psi_1$, and $\psi_2$, see Fig.~\ref{fig:2}(c). The white forbidden hexagons correspond to the configurations of $(\theta_{01},\theta_{12},\theta_{20})$ for which the closure of the triangle $r_{01}e^{i\theta_{01}}+r_{12}e^{i\theta_{12}}+r_{20}e^{i\theta_{20}}=0$ cannot be satisfied for any set of positive edge lengths $r_{ij}\equiv |\psi_i - \psi_j|>0$, see Appendix~\ref{app:opspace}. The boundaries of the hexagonal regions are also forbidden since they correspond to points where $\psi_i=\psi_j$ for some $i\neq j$, and thus do not belong in $\mc C_3(\mb C)$. In the allowed red (blue) regions, the above closure can be satisfied and the triangle spanned by  $\psi_0,\psi_1$ and $\psi_2$ has positive (negative) chirality $\chi=\text{sign}\left[\sin(\theta_{01}-\theta_{20})\right]$. At the nodal lines with $\chi=0$, indicated by dots in Fig.~\ref{fig:2}(a), the triangles collapse to a collinear configuration in the complex plane. For the gray, red, and blue nodal lines, $\psi_0$, $\psi_1$, and $\psi_2$ are located in the center, respectively. Hence, the gray node represents the ground state, while the red and blue nodes are energetically penalized by $U_0$. Finally, the primitive vectors of the space can be expressed as
$\vb{a}=2\pi(1,0,0)$, $\vb{b}=2\pi(0,0,1)$, and $\vb{c}=2\pi(1,1,1)$, such that any vector $n \vb{a}+m\vb{b}+l \vb{c}$ with $n,l,m\in\mathbb{Z}$ connects equivalent states $\bs\Psi$. This periodicity is reflected in Fig.~\ref{fig:2}(a) by coloring equivalent states on the nodal lines of the unit cell by the same color. 

\tt{Vortex classification}For a vortex contained in the patch $\cal A$, the texture of $\bs{\Psi}(\*r)$ on the boundary $\partial \cal A$ can be mapped to a trajectory traversing the allowed regions of the order parameter space, see, for example, the trajectory from 1 to 7 in Fig.~\ref{fig:2}. While in the extended order parameter space of Fig.~\ref{fig:2}(b), the trajectory can start and end at any equivalent points, in the unit cell, see Fig.~\ref{fig:2}(a), the trajectory must be a closed loop.
A trajectory advancing along the $(1,1,1)$ direction increments all values of $\theta_{ij}$ by the same amount, corresponding to a rigid rotation of the triangle spanned by $\psi_0$, $\psi_1$, and $\psi_2$. Fully advancing by $\vb{c}$ connects equivalent configurations. For example, starting at a gray ground-state node with state $\bs{\Psi}_0$ and advancing by $\vb{c}$ corresponds to a Goldstone mode that winds up $\varphi$ by $2\pi$. Besides the overall twist of the texture, nontrivial braiding is achieved by traversing the plane orthogonal to $(1,1,1)$ through the red, gray, or blue nodal lines, where the chirality $\chi$ switches. 

The full topological information of a vortex can be distilled by continuously deforming the trajectory in order parameter space into its simplest representation: a reduced sequence of primitive vectors connecting gray nodes, e.g., $\left[l \*c, \eta_1 \*v_1,\eta_2\*v_2,\ldots,\eta_S\*v_S\right]$ for $l\in\mathbb{Z}$, $\*v_i\in \{\*a,\*b\}$, $\eta_i \in\{\pm 1\}$, and $i\in\{1,\ldots,S\}$~\footnote{By reduced sequence, we mean that  consecutive entries of $\vb{v}$ and their inverses  $-\vb{v}$ with $\vb{v}\in\{\vb{a},\vb{b}\}$ have been removed, such that the sequence has the minimal number of primitive vectors.}. It is important to note that a trajectory advancing precisely along $\*a$ and $\*b$ as in Fig.~\ref{fig:2}(b) is impossible since the sides of the hexagon are forbidden. Hence, the reduced sequence must be understood as an \textit{asymptotic} limit of deforming the trajectory. Crucially, $\*a$ and $\*b$ do not commute due to the hexagonal holes in the order parameter space. In contrast, $\*{c}$ does commute with $\*a$ and $\*b$ since the space is invariant along $(1,1,1)$. With this in mind, any vortex can  be interpreted as the combination of an abelian vortex $[l\*c]$  together with a nonabelian vortex characterized by $\left[ \eta_1 \*v_1,\eta_2\*v_2,\ldots,\eta_S\*v_S\right]$~\footnote{In this description, a vortex with only relative winding $n_{12}=1$ is the composite object $[\*c,-\*a,-\*b]$, which is the pairing of an abelian vortex $[\vb{c}]$ with $[-\vb{a},-\vb{b}]$.}. Here, we have established the convention that the square brackets $[\ldots]$ signify equivalence under cyclic permutations of the sequence, such that the topological classification is independent of the choice of basepoint on $\partial\mc A$. The phase windings from Eq.~\eqref{eq:relwindingnumbers} of these vortices are given by  $(n_{01},n_{12},n_{20})=(l,l,l)$ for the abelian vortex $[l\*c]$ and $(n_{01},n_{12},n_{20})=(n,0,m)$ for the nonabelian vortex $\left[ \eta_1 \*v_1,\eta_2\*v_2,\ldots,\eta_S\*v_S\right]$, where $n$ and $m$ are the  net number of translations along $\*a$ and $\*b$, respectively. The elementary nonabelian vortices $[\*a]$ and $[\*b]$ carry a fraction of the abelian winding charge of $[\*c]$ and hence can be understood as fractional vortices. We remark that the abelian vortex $[\*c]$ corresponds to the generator of the center of the fundamental homotopy group $P_3$, $Z(P_3)=\mathbb Z$, while  the elementary nonabelian vortices $[\*a]$ and $[\*b]$ correspond to the generators of $P_3/Z(P_3)\cong F_2$, see Appendix~\ref{app:grouptheory}.

As an example, Fig.~\ref{fig:2} shows the trajectory (from 1 to 7) as we encircle a $\left[\*b,\*a\right]$ vortex.  In Fig.~\ref{fig:2}(b), starting at a gray node, the trajectory makes a half circle counterclockwise around a hexagonal hole, traversing through the blue and red nodes before coming back to a gray node. Under continuous transformation, this trajectory can be deformed into the asymptotic one that first advances along $\vb{b}$ and then along $\vb{a}$.  In Fig.~\ref{fig:2}(a), we see in the 3D  unit cell of the order parameter space that, in addition to traversing the plane, the braid trajectory must necessarily traverse in the $\vb{c}$ direction, since $\vb{a}$ and $\vb{b}$ are not orthogonal to $\vb{c}$. 
In  Fig.~\ref{fig:2}(c), we show the respective real-space texture of $\psi_0,\psi_1$ and $\psi_2$ on the loop $\partial \cal A$ with $0 \le  \ell< L$ enclosing the vortex.

Now, we are equipped to understand the 2D real-space structure of a nonabelian vortex $\left[ \eta_1 \*v_1,\eta_2\*v_2,\ldots,\eta_S\*v_S\right]$ such as the one shown within the black dashed circle of Fig.~\ref{fig:1}. For this, we notice that each primitive vector $\vb{a}$ and $\*b$ in the reduced sequence indicate crossings red and blue zero chirality nodes, respectively, see Fig.~\ref{fig:2}(b). Since these crossings are deviations from the gray ground-state node, they are penalized by repulsion $U_0$, and the characteristic length scale of such a texture variation is  $\xi \sim \sqrt{A/U_0}$. Hence, in real space, a vector $\vb{a}$ ($\vb{b}$) signifies a red (blue) string of width $\xi $ attached to the vortex core. Note that these strings maintain a sense of orientation depending on whether the red and blue nodes are crossed as we traverse from positive to negative chirality $\chi$ or vice versa (indicated by the light red and light blue shaded regions). The total number $S$ of $\*a$ and $\*b$  primitive vectors in  a cyclically reduced sequence corresponds to the total number of strings.
Crucially, since $\*a$ and $\*b$  do not commute, the red and blue strings cannot move past one another without creating new singularities. 
This endows these composite nonabelian vortices with exponential topological information. 
In particular, for a given number $S$, there are  $N_S\sim 3^S/S$ topologically distinct  nonabelian vortices for $S\gg 1$~\cite{coornaertIJAC05}, making them ideal carriers of topologically robust information~\footnote{Reference~\cite{ramsayJCA25} provides an exact expression for $N_S$, making use of Euler's totient function.}.

\tt{Brunnian vortices}To form a generic vortex, it is sufficient that upon taking a loop $\partial\mc A$ around the vortex in real space, the $\bs \Psi$ texture traces out a trajectory in order parameter space that starts and ends at identified points. However, vortices that correspond to a \textit{closed} loop in order parameter space [see dashed line in Fig.~\ref{fig:2}(b)] belong to an  intriguing class of defects, which we term Brunnian vortices~\footnote{The term Brunnian was chosen because the complex fields $\psi_0,\psi_1$ and $\psi_2$ of such textures trace out a 3-component Brunnian link in the space $\mb C\times S^1$.}. Remarkably, such defects all have zero winding number $n_{01}=n_{12}=n_{20}=0$. To detect their presence, we must introduce a second-order winding number~\cite{bergerJPA91} 
\begin{align}\label{eq:secondwinding}
    {\cal Q}^{(2)}&= \int_{\partial \cal A} \mathrm{d} {\ell}  \left(\partial_\ell \bs{\theta} \right) \cdot \vb{A}(\bs \theta)
\in \mathbb{Z},
\end{align}
where $\boldsymbol\theta=(\theta_{01},\theta_{12},\theta_{20})$ is parametrized on  a loop $\partial \cal A$ encircling the vortex  with  $0\le\ell<L$ and $0$ and $L$ are identified, $\bs{\theta}(0)=\bs{\theta}(L)$.  Here, $\vb{A}$ with $\vb{A}\cdot \vb{c}=0$ is a flat connection in order parameter space that is sensitive to the hexagonal holes in the plane orthogonal to $\vb{c}$,  see Appendix~\ref{app:secondorderwinding} for details. The number ${\cal Q}^{(2)}$ measures the net number of hexagons enclosed counterclockwise by the trajectory. For example, the simplest Brunnian vortex $[\*b,\*a,-\*b, -\*a]$ (the Borromean vortex~\cite{cromwell1998borromean}) encircles one hexagon counterclockwise, for which the second-order winding outputs ${\cal Q}^{(2)}=1$. This vortex is depicted within the black dashed circle of Fig.~\ref{fig:1}. For a given number of strings $S$, there are $\sim 3^S/S^2$ different Brunnian vortices for $S\gg1$, since only a fraction $\sim 1/S$ of all $\sim 3^S/S$ vortices correspond to a recurrent trajectory in order parameter space~\cite{kemptonARX16}.  Hence, they also carry exponential topological information. The single integer $\mc Q^{(2)}$, therefore, does not provide a faithful  classification of all Brunnian vortices~\footnote{As a concrete example, a trajectory which encloses one hexagon clockwise and a neighboring one counterclockwise is associated with a Brunnian vortex, yet $\mc Q^{(2)} = 0$.}. Rather, if $n_{ij}=0$ while  $\mc Q^{(2)}\neq 0$, then Brunnian vortices must be present. 

As a consequence of not carrying any winding numbers $n_{ij}$, these objects feel zero net Magnus forces in response to an external drive~\cite{soninPRB97}. However, a coupling which generically couples to texture will still give rise to a featureless drag of the topological texture. Moreover, they carry entropy so their signatures are  expected in thermal effects, for example in physics  similar to Berezinskii–Kosterlitz–Thouless transitions~\cite{kobayashiPRL19}.

\begin{figure}
    \centering  
    \includegraphics[width=\linewidth]{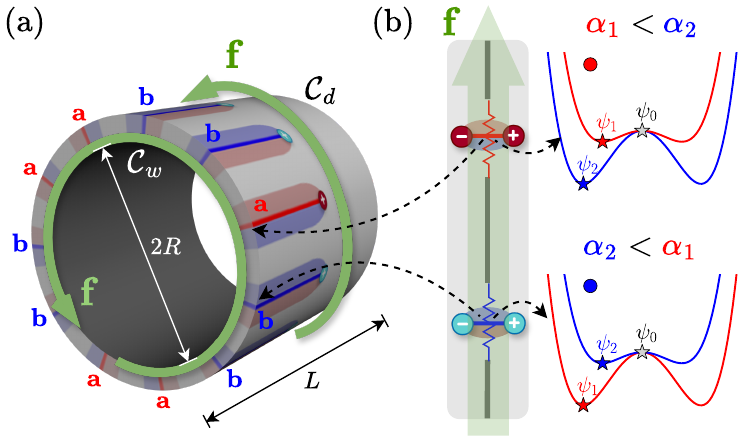}
    \caption{(a) Memory storage concept and (b) schematic of weak links facilitating vortex nucleation. The Mexican hat potentials show how $\alpha_1$ and $\alpha_2$ must be tuned to select for $[\*a]$ and $[\*b]$ vortices. The green arrows represent the external drive $\*f$, which nucleates the vortex-antivortex pairs at the weak links (red and blue zigzags) in (b) and transports the vortices from $\mc C_w$ to $\mc C_d$ via the Magnus force in (a).}
    \label{fig:cylinder}
\end{figure}

\tt{Nonabelian memory storage}To illustrate a potential technological application that takes advantage of nonabelian vortices, we propose a memory storage concept. For this, we assume an open-ended cylinder depicted in Fig.~\ref{fig:cylinder}(a), which has length $L$ and radius $R$. The goal is to store information on the cylinder by winding up the $\bs \Psi$ texture, specifically, by creating $S$ distinct red and blue strings extending along the cylinder axis. Since red and blue lines cannot cross,  the information is topologically protected.

Information can be written on the device by locally engineering weak points placed at locations $\*r_i$ for $i=1,\ldots,S$ along the leftmost writing edge $\mc C_w$ of the cylinder. At these points, vortex-antivortex pairs can be selectively nucleated~\cite{lesserPRB26} to encode topological information. For nucleation, we assume that we can control the Mexican hat potential of the condensates individually on the writing edge $\mc C_w$. Setting $\alpha_1<\alpha_2$ ($\alpha_2<\alpha_1$) facilitates the creation of an $[\*a]$-type ($[\*b]$-type) vortex-antivortex pair connected by a red (blue) string.

For information transmission, we bias these pairs so that the $[\*a]$ and $[\*b]$ vortices are sent along the axis of the cylinder towards the detecting edge $\mc C_d$, while the $[-\*a]$ and $[-\*b]$ antivortices are biased oppositely. To do so, we follow previous works on topological charge transport in abelian systems as a conceptual blueprint~\cite{daoPRB25,daoPRB26,jonesPRB20,zouPRB19,zouPRL20}. Namely, we can generally use an external drive $\*f$ to linearly couple to the winding via a term $-\vb{f}\cdot \left(\grad \theta_{01}+\grad \theta_{20}\right)$ in the free energy. For example, superfluid winding couples to rotations or curvature~\cite{satoRPP11}, and superconducting winding couples to the magnetic vector potential~\cite{coleman15}~\footnote{To couple to the relative winding $\grad \theta_{12}$, we would require an interaction of the Lifshitz form, which breaks inversion symmetry and has been discussed in the context of multiband superconductors~\cite{nagashimaPRB25}.}. Furthermore, winding in easy-plane magnets can couple to  electric currents~\cite{daoPRB25,zouPRB19,jonesPRB20} or in-plane electric fields~\cite{zhuPRB25}.  
Generically, such a coupling to winding generates a Magnus force on vortices, where the vortex flux is orthogonal to the applied drive~\cite{soninPRB97}. By applying a drive in the azimuthal direction, the vortices move along the cylinder axis. Since the vortices are dressed with topological strings, this scheme allows the transport of nonabelian information. Once the singular vortex cores have moved past the edges, the topological information is encoded by  the red and blue strings of the nonsingular $\bs{\Psi}$ texture. 

The geometry of a cylinder has two benefits. First, the hole allows us to avoid the energetic costs of the singular vortex cores that are created at the writing edge $\mc C_w$. Second, the energetic cost of imprinting the cylinder with the $\bs \Psi$ texture as the vortices traverse the bulk scales linearly with the length $L$ of the cylinder. This energetic cost is supplied by the constant surface drive $\*f$ circulating the cylinder. Once the singular vortex cores pass through the detector edge $\mc C_d$, the texture in the bulk and on the edges has no singularities. The system becomes metastable since relaxation of the texture would require nucleation of an energetically costly vortex core. Hence, the texture persists even if the drive $\*f$ is turned off, analogous to how a $\t U(1)$ superfluid can circulate indefinitely in a closed loop in the absence of any phase slip events. Since the information of the vortices can be stored without needing further power input, our device is a realization of nonvolatile memory storage. The amount of information that can be written on the device grows with the number of strings $S$ extending from $\mc C_w$ to $\mc C_d$. Without supplying more power after writing, we can estimate $S\sim R/\xi$, where $\xi\sim\sqrt{A/U_0}$ is the size of the chiral domains. The maximum number of distinct configurations is given by $2^S/S$ for $S\gg1$.

\tt{Discussion}Experimental realizations of nonabelian vortices in condensed matter systems remain elusive, with only a handful of genuine examples. Some prominent cases include liquid crystals with biaxial order~\cite{taylorPRL70, wuPRX25, merminRMP79} and the recently discovered magnetic phases of spinor Bose-Einstein condensates~\cite{xiaoNATC22,borghPRL16,semenoffPRL07}. Even in these examples, the nonabelian information encoded are classified by a finite group~\footnote{To be precise, the truly nonabelian structure is determined by the quotient of the fundamental group with its (abelian) center. In this work, we have that $\pi_1(\mc M)/Z[\pi_1(\mc M)] = P_3/Z(P_3) = F_2$, which is an infinite-sized group.}; hence, they do not have the same capacity for information storage as those discussed here. While we invoke a minimal model of a 2-component condensate, it is still an open question what would be an ideal material platform.  Moreover, our work blueprints how to treat vortices of the $\mc C_3(\mb C)$ order parameter space, even if the physical underpinnings are different. For instance, three-component superconductors with mutual Josephson couplings between the condensates can realize the order parameter space without relying on the Mexican hat potential, and are promising for future studies~\cite{rybakovPRI26,garaudPRB13,zhengSCI26}. Two-sublattice ferrimagnets are also an attractive candidate because the magnetic interactions can be tailored via material composition and heterostructure engineering to implement the $\mc C_3(\mb C)$ order parameter 
while conventional spintronic controls can electrically drive magnetic textures~\cite{kimNATM22,quessabSCIR20,zouPRR23, rybakovARX22}. 

Another intriguing direction would be to investigate the thermodynamic behavior of the proposed system. The red and blue strings connecting vortex-antivortex pairs are energetically costly and effectively confine the nonabelian vortices. However, at finite temperatures, the configurational entropy gained by producing the strings competes with the energetic cost, which can suppress the effective tension of the strings and lead to proliferation of strings and vortex-antivortex pairs. This suggests the possibility for a thermally driven phase transition, analogous to the proliferation of strings and deconfinement of monopoles in quantum spin-ice~\cite{wanPRL12,panNATP16}.

\smallskip

\begin{acknowledgments}
\tt{Acknowledgments}We thank Shane P. Kelly, Shu Zhang, Yanyan Zhu, and Filipp N. Rybakov for insightful discussions.  This work is supported by the U.S. Department of Energy, Office of Basic Energy Sciences, under Grant No. DE-SC0012190.
\end{acknowledgments}

\setcounter{secnumdepth}{2}
\appendix
\section{Order parameter space}\label{app:opspace}
We construct the reduced order parameter space spanned by the three relative angles $(\theta_{01},\theta_{12},\theta_{20})$, which is shown in Fig.~\ref{fig:2}(a,b) and formally follows from the image of the map in Eq.~\eqref{eq:relangles}~\cite{bergerJPA91}. To this end, we define $\psi_i - \psi_j=r_{ij}e^{i\theta_{ij}}$ with $r_{ij}\equiv |\psi_i - \psi_j|>0$ and $\theta_{ij}=\arg(\psi_i-\psi_j)$,  so that we can formulate the closure relation
\begin{align}\label{eq:closure}
r_{01}e^{i\theta_{01}}+r_{12}e^{i\theta_{12}}+r_{20}e^{i\theta_{20}}=0.
\end{align}
We rewrite this condition via $\vb{M} \vb{r}=0$ with $\vb{r}= (r_{01},r_{12},r_{20})$ and the matrix
\begin{align}
\vb{M}=\begin{pmatrix}
\cos\theta_{01} & \cos\theta_{12} & \cos\theta_{20} \\
\sin\theta_{01} & \sin\theta_{12} & \sin\theta_{20}
\end{pmatrix},
\end{align}
where the first and second rows follow from the real and imaginary parts of Eq.~\eqref{eq:closure}, respectively. The unique solution of $\vb{M}\vb{r}=0$ is given by the null vector
\begin{align}
\vb{r}= d 
\begin{pmatrix}
\sin(\theta_{12}-\theta_{20})\\
\sin(\theta_{20}-\theta_{01})\\
\sin(\theta_{01}-\theta_{12})
\end{pmatrix}.
\end{align}
To ensure $\vb{r}>\vb{0}$, we distinguish between solutions for which $d>0$ and $d<0$ with
\begin{align}\label{eq:regions}
 \begin{pmatrix}
\sin(\theta_{12}-\theta_{20})\\
\sin(\theta_{20}-\theta_{01})\\
\sin(\theta_{01}-\theta_{12})
\end{pmatrix}>\vb{0},  \quad \begin{pmatrix}
\sin(\theta_{12}-\theta_{20})\\
\sin(\theta_{20}-\theta_{01})\\
\sin(\theta_{01}-\theta_{12})
\end{pmatrix}<\vb{0},
\end{align}
respectively. 
The former gives rise to a region of negative chirality (shaded light blue), while the latter gives rise to a region of positive chirality (shaded light red), where the chirality is $\chi=\text{sign}\left[\sin(\theta_{01}-\theta_{20})\right]$.
Note that shifting all angles $\theta_{ij}$ by a constant does not change the inequalities in Eq.~\eqref{eq:regions}, which establishes the translational invariance of the order parameter space along the $(1,1,1)$ direction. 

\section{Group-theoric classification}\label{app:grouptheory}
\begin{figure}
    \centering
    \includegraphics[width=0.6\linewidth]{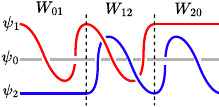}
    \caption{Depiction of the braid element $W_{01}W_{12}W_{20}$, which is also the generator of the center $Z(P_3) = \mb Z$ and describes the full twist of the braid.}
    \label{fig:twist}
\end{figure}

The starting point for a group theoretic analysis is to inspect the fundamental homotopy group of the order parameter space $\mathcal C_3(\mb C)$. Taking a closed loop $\partial \cal A$ that encircles the vortices on patch $\cal A$, the topological information of the texture of $\partial\mc A$ is encoded by the edge homotopy 
\begin{equation}\label{eq:edgehomotopy}
\pi_1[\mathcal C_3(\mb C)] = P_3 = \langle W_{01},W_{12}, W_{20}\rangle.
\end{equation}
This is the pure braid group generated by the group elements $W_{01}$, $W_{12}$, and $W_{20}$, which must satisfy the cyclic group relations~\cite{leeJCM10}
\begin{equation}
    W_{01}W_{12}W_{20} = W_{20}W_{01}W_{12} = W_{12}W_{20}W_{01},
\end{equation}
ensuring the pure-braid analog of the Yang-Baxter relation for three strands. The generators $W_{ij}$ can be identified with a braiding operation that winds $\psi_i$ around $\psi_j$ by $2\pi$, while introducing no other windings, see Fig.~\ref{fig:twist}. For a group element of $ P_3$, the  winding number $n_{ij}$ in the set $(n_{01},n_{12},n_{20})$ from Eq.~\eqref{eq:relwindingnumbers} simply counts the ``net"  number of $W_{ij}$'s in the group element, meaning $n_{ij} = \#W_{ij} - \#W_{ij}^{-1}$. Using group theoretic jargon, these three independent winding numbers correspond to the abelianization of the fundamental homotopy, $\pi_1^\mathrm{ab}\left[\mathcal C_3(\mb C)\right]=\mathbb Z^3$.

While these abelianized winding numbers $n_{ij}$ contain partial information about the enclosed vortices, the noncommutativity of the $W_{ij}$ generators suggests that a nonabelian vortex cannot simply be decomposed into three types of abelian vortices with winding numbers $n_{ij}$. Rather, we propose to identify the abelian vortices of the order parameter space  with the center of the fundamental homotopy group, $Z(P_3)= \langle c\rangle \cong\mb Z$, where the generator $c=W_{01}W_{12}W_{20}$ is a full twist of the braid which increments each of the relative winding numbers by $+1$.

The truly nonabelian information is encoded by the group quotient $P_3/ Z(P_3) \cong F_2$, where $F_2=\langle a,b\rangle$ is the free group with generators $a$ and $b$, where we identify $a=W_{01}$ and $b=W_{20}$. In fact, here, $P_3$ can be factored into the product of groups $P_3 \cong \mb Z\times F_2$. Since the classification of an associated nonabelian vortex should be independent of any basepoint on $\partial \cal A$, they are characterized by cyclic permutations of $g_1g_2\ldots g_n$ with $g_i\in\{a,b,a^{-1},b^{-1}\}$. For example, $aba^{-1}\sim a^{-1}a b= b$. This equivalence relation is exactly captured by the conjugacy classes of $F_2$, which are formally defined such that $g,g'\in F_2$ are conjugate if there exists an $h\in F_2$ such that $g' = hgh^{-1}$. Note that the generators $a$ and $b$ as well as the center element $c$ correspond to the primitive vectors $\vb{a}$, $\vb{b}$, and $\vb{c}$ defined above. 

We remark that within the vortex cores, the order parameter $\bs{\Psi}$ violates $\psi_i\neq\psi_j$ and, therefore, leaves the order parameter space $\mathcal C_3(\mb C)$. Formally, the topology of such a 2D vortex texture on a patch $\cal A$ can be captured by the relative homotopy~\cite{rybakovPRB25}
\begin{align}
    \pi_2[\mb C^3,\mathcal C_3(\mb C)]=P_3,
\end{align}
which allows the complex fields $\psi_0$, $\psi_1$, and $\psi_2$ in the interior of a patch $\cal A$  to take any value in $\mb C^3$, while its boundary  $\partial\cal A$  is restricted to $\mathcal C_3(\mb C)$. Here, this relative homotopy group outputs the same braid group $P_3$ as the edge homotopy of Eq.~\eqref{eq:edgehomotopy}, which indicates that the edge $\partial \cal A$ is sufficient to classify vortex textures. In general, however, this must not be the case if true bulk defects are present.   

\section{Second-order winding number}\label{app:secondorderwinding}
To obtain the second-order winding number, we define the connection in the reduced order parameter space spanned by the three relative angles $\boldsymbol\theta=(\theta_{01},\theta_{12},\theta_{20})$ via~\cite{bergerJPA91}
\begin{align}
    \begin{split}
    \vb{A}=&\Big[\left(\lambda_{01}-\lambda_{20}\right)\grad_{\vb{\theta}}\lambda_{12}+\left(\lambda_{12}-\lambda_{01}\right) \grad_{\vb{\theta}}\lambda_{20}\\  &+\left(\lambda_{20} -\lambda_{12}\right)\grad_{\vb{\theta}}\lambda_{01}\Big]/2,
    \end{split}
\end{align}
where $\lambda_{ij}=\,\ln(\psi_i-\psi_j)/2\pi i$ and $\grad_{\vb{\theta}}=(\partial_{\theta_{01}},\partial_{\theta_{12}},\partial_{\theta_{20}})$. Here, we use the parametrization $\psi_0=0$, and $\psi_{j}=\sqrt{\rho_{j}}e^{i\theta_{j}}$ with $j\in\{1,2\}$. The condensate phases are given by $\theta_1=\theta_{01}+\pi$ and $\theta_2=\theta_{20}$. In addition, $\theta_{12}=\arg(\psi_1-\psi_2)$ fixes the ratio $\sqrt{\rho_2/\rho_1}=\sin(\theta_{01}-\theta_{12})/\sin(\theta_{12}-\theta_{20})$. 
Note that the connection is defined such that it is invariant under a common scaling and translation of the fields $\psi_i$.  
Now, it is straightforward to show that 
\begin{align}
\begin{split}
  \grad_{\vb{\theta}}\times \vb{A}=&\grad_{\vb{\theta}}\lambda_{01}\times \grad_{\vb{\theta}}\lambda_{12}+\grad_{\vb{\theta}}\lambda_{12}\times \grad_{\vb{\theta}}\lambda_{20}\\ &+\grad_{\vb{\theta}}\lambda_{20}\times \grad_{\vb{\theta}}\lambda_{01}=0,
\end{split}
\end{align}
which holds in the allowed regions of the order parameter space, away from the edges of the forbidden hexagons on which $\psi_i=\psi_j$ and hence, $\lambda_{ij}$ becomes singular. The relation is known as the Arnold relation~\cite{vladimir1969arnol}. 
Since $\grad_{\vb{\theta}}\times \vb{A}=0$,  the connection $\vb{A}$ is flat (no curvature) and can be used to detect the topology of the order parameter space by parallel transport along noncontractible loops $\cal C$ (referred to as holonomy). Since $\vb{A}\cdot \vb{c}=0$, it is sufficient to consider loops $\cal C$ projected to the plane orthogonal to $\vb{c}$, see Fig.~\ref{fig:2}(b). 
In fact, for any closed loop, we obtain an integer
\begin{align}
   {\cal Q}^{(2)}= \oint_{\cal C} \mathrm{d}\boldsymbol{\theta}\cdot \vb{A} \in \mathbb{Z}.
\end{align}
Here, since the connection is flat, any loop $\cal C$ can be deformed into a sum of elementary integrals enclosing  single hexagons without changing the value of ${\cal Q}^{(2)}$. One can (e.g.  numerically) show that the counterclockwise (clockwise) integration around a single hexagon gives $+1$ ($-1$), see the dashed circle in Fig.~\ref{fig:2}(b). Hence, ${\cal Q}^{(2)}$ measures the net number of holes encircled in a counterclockwise fashion. To obtain Eq.~\eqref{eq:secondwinding}, we  parametrize  on a loop $\partial \cal A$ in real space using $\mathrm{d}\boldsymbol\theta =\mathrm{d}{\ell}\cdot \partial_{\ell} \bs{\theta} $.

\bibliography{mybib.bib}

\providecommand{\noopsort}[1]{}\providecommand{\singleletter}[1]{#1}%
\begin{thebibliography}{55}%
\makeatletter
\providecommand \@ifxundefined [1]{%
 \@ifx{#1\undefined}
}%
\providecommand \@ifnum [1]{%
 \ifnum #1\expandafter \@firstoftwo
 \else \expandafter \@secondoftwo
 \fi
}%
\providecommand \@ifx [1]{%
 \ifx #1\expandafter \@firstoftwo
 \else \expandafter \@secondoftwo
 \fi
}%
\providecommand \natexlab [1]{#1}%
\providecommand \enquote  [1]{``#1''}%
\providecommand \bibnamefont  [1]{#1}%
\providecommand \bibfnamefont [1]{#1}%
\providecommand \citenamefont [1]{#1}%
\providecommand \href@noop [0]{\@secondoftwo}%
\providecommand \href [0]{\begingroup \@sanitize@url \@href}%
\providecommand \@href[1]{\@@startlink{#1}\@@href}%
\providecommand \@@href[1]{\endgroup#1\@@endlink}%
\providecommand \@sanitize@url [0]{\catcode `\\12\catcode `\$12\catcode `\&12\catcode `\#12\catcode `\^12\catcode `\_12\catcode `\%12\relax}%
\providecommand \@@startlink[1]{}%
\providecommand \@@endlink[0]{}%
\providecommand \url  [0]{\begingroup\@sanitize@url \@url }%
\providecommand \@url [1]{\endgroup\@href {#1}{\urlprefix }}%
\providecommand \urlprefix  [0]{URL }%
\providecommand \Eprint [0]{\href }%
\providecommand \doibase [0]{https://doi.org/}%
\providecommand \selectlanguage [0]{\@gobble}%
\providecommand \bibinfo  [0]{\@secondoftwo}%
\providecommand \bibfield  [0]{\@secondoftwo}%
\providecommand \translation [1]{[#1]}%
\providecommand \BibitemOpen [0]{}%
\providecommand \bibitemStop [0]{}%
\providecommand \bibitemNoStop [0]{.\EOS\space}%
\providecommand \EOS [0]{\spacefactor3000\relax}%
\providecommand \BibitemShut  [1]{\csname bibitem#1\endcsname}%
\let\auto@bib@innerbib\@empty
\bibitem [{\citenamefont {Mermin}(1979)}]{merminRMP79}%
  \BibitemOpen
  \bibfield  {author} {\bibinfo {author} {\bibfnamefont {N.~D.}\ \bibnamefont {Mermin}},\ }\bibfield  {title} {\bibinfo {title} {The topological theory of defects in ordered media},\ }\href {https://link.aps.org/doi/10.1103/RevModPhys.51.591} {\bibfield  {journal} {\bibinfo  {journal} {Rev. Mod. Phys.}\ }\textbf {\bibinfo {volume} {51}},\ \bibinfo {pages} {591} (\bibinfo {year} {1979})}\BibitemShut {NoStop}%
\bibitem [{\citenamefont {Wu}\ \emph {et~al.}(2025)\citenamefont {Wu}, \citenamefont {Valenzuela}, \citenamefont {Bowick},\ and\ \citenamefont {Smalyukh}}]{wuPRX25}%
  \BibitemOpen
  \bibfield  {author} {\bibinfo {author} {\bibfnamefont {J.-S.}\ \bibnamefont {Wu}}, \bibinfo {author} {\bibfnamefont {R.~A.}\ \bibnamefont {Valenzuela}}, \bibinfo {author} {\bibfnamefont {M.~J.}\ \bibnamefont {Bowick}},\ and\ \bibinfo {author} {\bibfnamefont {I.~I.}\ \bibnamefont {Smalyukh}},\ }\bibfield  {title} {\bibinfo {title} {Topological rigidity and non-abelian defect junctions in chiral nematic systems with effective biaxial symmetry},\ }\href {https://link.aps.org/doi/10.1103/PhysRevX.15.021036} {\bibfield  {journal} {\bibinfo  {journal} {Phys. Rev. X}\ }\textbf {\bibinfo {volume} {15}},\ \bibinfo {pages} {021036} (\bibinfo {year} {2025})}\BibitemShut {NoStop}%
\bibitem [{\citenamefont {Tserkovnyak}(2018)}]{tserkovnyakJAP18}%
  \BibitemOpen
  \bibfield  {author} {\bibinfo {author} {\bibfnamefont {Y.}~\bibnamefont {Tserkovnyak}},\ }\bibfield  {title} {\bibinfo {title} {Perspective: (beyond) spin transport in insulators},\ }\href {https://doi.org/10.1063/1.5054123} {\bibfield  {journal} {\bibinfo  {journal} {J. Appl. Phys.}\ }\textbf {\bibinfo {volume} {124}},\ \bibinfo {pages} {190901} (\bibinfo {year} {2018})}\BibitemShut {NoStop}%
\bibitem [{\citenamefont {Zou}\ \emph {et~al.}(2019)\citenamefont {Zou}, \citenamefont {Kim},\ and\ \citenamefont {Tserkovnyak}}]{zouPRB19}%
  \BibitemOpen
  \bibfield  {author} {\bibinfo {author} {\bibfnamefont {J.}~\bibnamefont {Zou}}, \bibinfo {author} {\bibfnamefont {S.~K.}\ \bibnamefont {Kim}},\ and\ \bibinfo {author} {\bibfnamefont {Y.}~\bibnamefont {Tserkovnyak}},\ }\bibfield  {title} {\bibinfo {title} {Topological transport of vorticity in heisenberg magnets},\ }\href {https://link.aps.org/doi/10.1103/PhysRevB.99.180402} {\bibfield  {journal} {\bibinfo  {journal} {Phys. Rev. B}\ }\textbf {\bibinfo {volume} {99}},\ \bibinfo {pages} {180402(R)} (\bibinfo {year} {2019})}\BibitemShut {NoStop}%
\bibitem [{\citenamefont {Zou}\ \emph {et~al.}(2020)\citenamefont {Zou}, \citenamefont {Zhang},\ and\ \citenamefont {Tserkovnyak}}]{zouPRL20}%
  \BibitemOpen
  \bibfield  {author} {\bibinfo {author} {\bibfnamefont {J.}~\bibnamefont {Zou}}, \bibinfo {author} {\bibfnamefont {S.}~\bibnamefont {Zhang}},\ and\ \bibinfo {author} {\bibfnamefont {Y.}~\bibnamefont {Tserkovnyak}},\ }\bibfield  {title} {\bibinfo {title} {Topological transport of deconfined hedgehogs in magnets},\ }\href {https://link.aps.org/doi/10.1103/PhysRevLett.125.267201} {\bibfield  {journal} {\bibinfo  {journal} {Phys. Rev. Lett.}\ }\textbf {\bibinfo {volume} {125}},\ \bibinfo {pages} {267201} (\bibinfo {year} {2020})}\BibitemShut {NoStop}%
\bibitem [{\citenamefont {Dao}\ \emph {et~al.}(2025)\citenamefont {Dao}, \citenamefont {Zou}, \citenamefont {Kleinherbers},\ and\ \citenamefont {Tserkovnyak}}]{daoPRB25}%
  \BibitemOpen
  \bibfield  {author} {\bibinfo {author} {\bibfnamefont {C.}~\bibnamefont {Dao}}, \bibinfo {author} {\bibfnamefont {J.}~\bibnamefont {Zou}}, \bibinfo {author} {\bibfnamefont {E.}~\bibnamefont {Kleinherbers}},\ and\ \bibinfo {author} {\bibfnamefont {Y.}~\bibnamefont {Tserkovnyak}},\ }\bibfield  {title} {\bibinfo {title} {Topological transport of vorticity on curved magnetic membranes},\ }\href {https://link.aps.org/doi/10.1103/PhysRevB.111.L100401} {\bibfield  {journal} {\bibinfo  {journal} {Phys. Rev. B}\ }\textbf {\bibinfo {volume} {111}},\ \bibinfo {pages} {L100401} (\bibinfo {year} {2025})}\BibitemShut {NoStop}%
\bibitem [{\citenamefont {Dao}\ \emph {et~al.}(2026)\citenamefont {Dao}, \citenamefont {Kleinherbers}, \citenamefont {Brekke},\ and\ \citenamefont {Tserkovnyak}}]{daoPRB26}%
  \BibitemOpen
  \bibfield  {author} {\bibinfo {author} {\bibfnamefont {C.}~\bibnamefont {Dao}}, \bibinfo {author} {\bibfnamefont {E.}~\bibnamefont {Kleinherbers}}, \bibinfo {author} {\bibfnamefont {B.}~\bibnamefont {Brekke}},\ and\ \bibinfo {author} {\bibfnamefont {Y.}~\bibnamefont {Tserkovnyak}},\ }\bibfield  {title} {\bibinfo {title} {Topological hydrodynamics in ferromagnetic superconductors},\ }\href {https://link.aps.org/doi/10.1103/52jm-l96p} {\bibfield  {journal} {\bibinfo  {journal} {Phys. Rev. B}\ }\textbf {\bibinfo {volume} {113}},\ \bibinfo {pages} {174515} (\bibinfo {year} {2026})}\BibitemShut {NoStop}%
\bibitem [{\citenamefont {Rybakov}\ \emph {et~al.}(2025)\citenamefont {Rybakov}, \citenamefont {Eriksson},\ and\ \citenamefont {Kiselev}}]{rybakovPRB25}%
  \BibitemOpen
  \bibfield  {author} {\bibinfo {author} {\bibfnamefont {F.~N.}\ \bibnamefont {Rybakov}}, \bibinfo {author} {\bibfnamefont {O.}~\bibnamefont {Eriksson}},\ and\ \bibinfo {author} {\bibfnamefont {N.~S.}\ \bibnamefont {Kiselev}},\ }\bibfield  {title} {\bibinfo {title} {Topological invariants of vortices, merons, skyrmions, and their combinations in continuous and discrete systems},\ }\href {https://link.aps.org/doi/10.1103/PhysRevB.111.134417} {\bibfield  {journal} {\bibinfo  {journal} {Phys. Rev. B}\ }\textbf {\bibinfo {volume} {111}},\ \bibinfo {pages} {134417} (\bibinfo {year} {2025})}\BibitemShut {NoStop}%
\bibitem [{\citenamefont {Rybakov}\ and\ \citenamefont {Eriksson}(2022)}]{rybakovARX22}%
  \BibitemOpen
  \bibfield  {author} {\bibinfo {author} {\bibfnamefont {F.~N.}\ \bibnamefont {Rybakov}}\ and\ \bibinfo {author} {\bibfnamefont {O.}~\bibnamefont {Eriksson}},\ }\href {https://arxiv.org/abs/2205.15264} {\bibinfo {title} {Non-abelian vortices in magnets}} (\bibinfo {year} {2022}),\ \Eprint {https://arxiv.org/abs/2205.15264} {arXiv:2205.15264 [cond-mat.str-el]} \BibitemShut {NoStop}%
\bibitem [{Note1()}]{Note1}%
  \BibitemOpen
  \bibinfo {note} {In contrast, defects stemming from an order parameter consisting of only \protect \textit {two} ordered complex numbers are classified by the winding number, $\pi _1\left [ C_2(\protect \mathbb {C})\right ]=\protect \mathbb {Z}$, which simply counts the winding of $\psi _1$ around $\psi _0$ along a closed loop.}\BibitemShut {Stop}%
\bibitem [{\citenamefont {Li}\ \emph {et~al.}(2015)\citenamefont {Li}, \citenamefont {Liew}, \citenamefont {Egorov},\ and\ \citenamefont {Ostrovskaya}}]{liPRB15}%
  \BibitemOpen
  \bibfield  {author} {\bibinfo {author} {\bibfnamefont {G.}~\bibnamefont {Li}}, \bibinfo {author} {\bibfnamefont {T.~C.~H.}\ \bibnamefont {Liew}}, \bibinfo {author} {\bibfnamefont {O.~A.}\ \bibnamefont {Egorov}},\ and\ \bibinfo {author} {\bibfnamefont {E.~A.}\ \bibnamefont {Ostrovskaya}},\ }\bibfield  {title} {\bibinfo {title} {Incoherent excitation and switching of spin states in exciton-polariton condensates},\ }\href {https://link.aps.org/doi/10.1103/PhysRevB.92.064304} {\bibfield  {journal} {\bibinfo  {journal} {Phys. Rev. B}\ }\textbf {\bibinfo {volume} {92}},\ \bibinfo {pages} {064304} (\bibinfo {year} {2015})}\BibitemShut {NoStop}%
\bibitem [{\citenamefont {Son}\ and\ \citenamefont {Stephanov}(2002)}]{sonPRA02}%
  \BibitemOpen
  \bibfield  {author} {\bibinfo {author} {\bibfnamefont {D.~T.}\ \bibnamefont {Son}}\ and\ \bibinfo {author} {\bibfnamefont {M.~A.}\ \bibnamefont {Stephanov}},\ }\bibfield  {title} {\bibinfo {title} {Domain walls of relative phase in two-component bose-einstein condensates},\ }\href {https://link.aps.org/doi/10.1103/PhysRevA.65.063621} {\bibfield  {journal} {\bibinfo  {journal} {Phys. Rev. A}\ }\textbf {\bibinfo {volume} {65}},\ \bibinfo {pages} {063621} (\bibinfo {year} {2002})}\BibitemShut {NoStop}%
\bibitem [{\citenamefont {Kasamatsu}\ \emph {et~al.}(2004)\citenamefont {Kasamatsu}, \citenamefont {Tsubota},\ and\ \citenamefont {Ueda}}]{kasamtsuPRL04}%
  \BibitemOpen
  \bibfield  {author} {\bibinfo {author} {\bibfnamefont {K.}~\bibnamefont {Kasamatsu}}, \bibinfo {author} {\bibfnamefont {M.}~\bibnamefont {Tsubota}},\ and\ \bibinfo {author} {\bibfnamefont {M.}~\bibnamefont {Ueda}},\ }\bibfield  {title} {\bibinfo {title} {Vortex molecules in coherently coupled two-component bose-einstein condensates},\ }\href {https://link.aps.org/doi/10.1103/PhysRevLett.93.250406} {\bibfield  {journal} {\bibinfo  {journal} {Phys. Rev. Lett.}\ }\textbf {\bibinfo {volume} {93}},\ \bibinfo {pages} {250406} (\bibinfo {year} {2004})}\BibitemShut {NoStop}%
\bibitem [{\citenamefont {Dzyapko}\ \emph {et~al.}(2017)\citenamefont {Dzyapko}, \citenamefont {Lisenkov}, \citenamefont {Nowik-Boltyk}, \citenamefont {Demidov}, \citenamefont {Demokritov}, \citenamefont {Koene}, \citenamefont {Kirilyuk}, \citenamefont {Rasing}, \citenamefont {Tiberkevich},\ and\ \citenamefont {Slavin}}]{dzyapkoPRB17}%
  \BibitemOpen
  \bibfield  {author} {\bibinfo {author} {\bibfnamefont {O.}~\bibnamefont {Dzyapko}}, \bibinfo {author} {\bibfnamefont {I.}~\bibnamefont {Lisenkov}}, \bibinfo {author} {\bibfnamefont {P.}~\bibnamefont {Nowik-Boltyk}}, \bibinfo {author} {\bibfnamefont {V.~E.}\ \bibnamefont {Demidov}}, \bibinfo {author} {\bibfnamefont {S.~O.}\ \bibnamefont {Demokritov}}, \bibinfo {author} {\bibfnamefont {B.}~\bibnamefont {Koene}}, \bibinfo {author} {\bibfnamefont {A.}~\bibnamefont {Kirilyuk}}, \bibinfo {author} {\bibfnamefont {T.}~\bibnamefont {Rasing}}, \bibinfo {author} {\bibfnamefont {V.}~\bibnamefont {Tiberkevich}},\ and\ \bibinfo {author} {\bibfnamefont {A.}~\bibnamefont {Slavin}},\ }\bibfield  {title} {\bibinfo {title} {Magnon-magnon interactions in a room-temperature magnonic bose-einstein condensate},\ }\href {https://link.aps.org/doi/10.1103/PhysRevB.96.064438} {\bibfield  {journal} {\bibinfo  {journal} {Phys. Rev. B}\ }\textbf {\bibinfo {volume} {96}},\ \bibinfo {pages} {064438} (\bibinfo {year} {2017})}\BibitemShut
  {NoStop}%
\bibitem [{\citenamefont {Leggett}(1966)}]{leggettPTP66}%
  \BibitemOpen
  \bibfield  {author} {\bibinfo {author} {\bibfnamefont {A.~J.}\ \bibnamefont {Leggett}},\ }\bibfield  {title} {\bibinfo {title} {Number-phase fluctuations in two-band superconductors},\ }\href {https://doi.org/10.1143/PTP.36.901} {\bibfield  {journal} {\bibinfo  {journal} {Prog. Theor. Phys.}\ }\textbf {\bibinfo {volume} {36}},\ \bibinfo {pages} {901} (\bibinfo {year} {1966})}\BibitemShut {NoStop}%
\bibitem [{\citenamefont {Babaev}(2002)}]{babaevPRL02}%
  \BibitemOpen
  \bibfield  {author} {\bibinfo {author} {\bibfnamefont {E.}~\bibnamefont {Babaev}},\ }\bibfield  {title} {\bibinfo {title} {Vortices with fractional flux in two-gap superconductors and in extended faddeev model},\ }\href {https://link.aps.org/doi/10.1103/PhysRevLett.89.067001} {\bibfield  {journal} {\bibinfo  {journal} {Phys. Rev. Lett.}\ }\textbf {\bibinfo {volume} {89}},\ \bibinfo {pages} {067001} (\bibinfo {year} {2002})}\BibitemShut {NoStop}%
\bibitem [{\citenamefont {Babaev}\ \emph {et~al.}(2004)\citenamefont {Babaev}, \citenamefont {Sudb{\o}},\ and\ \citenamefont {Ashcroft}}]{babaevNAT09}%
  \BibitemOpen
  \bibfield  {author} {\bibinfo {author} {\bibfnamefont {E.}~\bibnamefont {Babaev}}, \bibinfo {author} {\bibfnamefont {A.}~\bibnamefont {Sudb{\o}}},\ and\ \bibinfo {author} {\bibfnamefont {N.~W.}\ \bibnamefont {Ashcroft}},\ }\bibfield  {title} {\bibinfo {title} {A superconductor to superfluid phase transition in liquid metallic hydrogen},\ }\href {https://doi.org/10.1038/nature02910} {\bibfield  {journal} {\bibinfo  {journal} {Nature}\ }\textbf {\bibinfo {volume} {431}},\ \bibinfo {pages} {666} (\bibinfo {year} {2004})}\BibitemShut {NoStop}%
\bibitem [{\citenamefont {Tanaka}(2001)}]{tanakaPRL01}%
  \BibitemOpen
  \bibfield  {author} {\bibinfo {author} {\bibfnamefont {Y.}~\bibnamefont {Tanaka}},\ }\bibfield  {title} {\bibinfo {title} {Soliton in two-band superconductor},\ }\href {https://link.aps.org/doi/10.1103/PhysRevLett.88.017002} {\bibfield  {journal} {\bibinfo  {journal} {Phys. Rev. Lett.}\ }\textbf {\bibinfo {volume} {88}},\ \bibinfo {pages} {017002} (\bibinfo {year} {2001})}\BibitemShut {NoStop}%
\bibitem [{Note2()}]{Note2}%
  \BibitemOpen
  \bibinfo {note} {It is interesting to note that upon turning off the $U(\psi _1,\psi _2)$ interaction, the $[\protect \mathbf {a}]$ and $[\protect \mathbf {b}]$ vortices revert to the conventional abelian $\protect \text {U}(1)$ vortices of the $\psi _1$ and $\psi _2$ condensates, respectively. Furthermore, $[\protect \mathbf {c}]$ reverts to its constituent condensate vortices, which are now decoupled, since the $n_{12}$ winding number is no longer protected.}\BibitemShut {Stop}%
\bibitem [{Note3()}]{Note3}%
  \BibitemOpen
  \bibinfo {note} {To better protect the order parameter space, a more nonlinear or localized repulsion can be used. For example, a flat rim of the Mexican hat potential paired with a sufficiently local repulsion that diverges for $\psi _1=\psi _2$ would be ideal.}\BibitemShut {Stop}%
\bibitem [{Note4()}]{Note4}%
  \BibitemOpen
  \bibinfo {note} {The configuration space $\protect \mathcal C_3(\protect \mathbb {C})$ can be understood as an effective order parameter space, in the sense that it parametrizes the low-energy manifold of the system. However, different configurations $\protect \boldsymbol {\Psi }$ do not generally have the same energy.}\BibitemShut {Stop}%
\bibitem [{Note5()}]{Note5}%
  \BibitemOpen
  \bibinfo {note} {By reduced sequence, we mean that consecutive entries of $\vb {v}$ and their inverses $-\vb {v}$ with $\vb {v}\in \{\vb {a},\vb {b}\}$ have been removed, such that the sequence has the minimal number of primitive vectors.}\BibitemShut {Stop}%
\bibitem [{Note6()}]{Note6}%
  \BibitemOpen
  \bibinfo {note} {In this description, a vortex with only relative winding $n_{12}=1$ is the composite object $[\protect \mathbf {c},-\protect \mathbf {a},-\protect \mathbf {b}]$, which is the pairing of an abelian vortex $[\vb {c}]$ with $[-\vb {a},-\vb {b}]$.}\BibitemShut {Stop}%
\bibitem [{\citenamefont {Coornaert}(2005)}]{coornaertIJAC05}%
  \BibitemOpen
  \bibfield  {author} {\bibinfo {author} {\bibfnamefont {M.}~\bibnamefont {Coornaert}},\ }\bibfield  {title} {\bibinfo {title} {Asymptotic growth of conjugacy classes in finitely-generated free groups},\ }\href {https://doi.org/10.1142/S0218196705002505} {\bibfield  {journal} {\bibinfo  {journal} {Int. J. Algebra Comput.}\ }\textbf {\bibinfo {volume} {15}},\ \bibinfo {pages} {887} (\bibinfo {year} {2005})}\BibitemShut {NoStop}%
\bibitem [{Note7()}]{Note7}%
  \BibitemOpen
  \bibinfo {note} {Reference~\cite {ramsayJCA25} provides an exact expression for $N_S$, making use of Euler's totient function.}\BibitemShut {Stop}%
\bibitem [{Note8()}]{Note8}%
  \BibitemOpen
  \bibinfo {note} {The term Brunnian was chosen because the complex fields $\psi _0,\psi _1$ and $\psi _2$ of such textures trace out a 3-component Brunnian link in the space $\protect \mathbb {C}\times S^1$.}\BibitemShut {Stop}%
\bibitem [{\citenamefont {Berger}(1991)}]{bergerJPA91}%
  \BibitemOpen
  \bibfield  {author} {\bibinfo {author} {\bibfnamefont {M.~A.}\ \bibnamefont {Berger}},\ }\bibfield  {title} {\bibinfo {title} {Third-order braid invariants},\ }\href {https://doi.org/10.1088/0305-4470/24/17/019} {\bibfield  {journal} {\bibinfo  {journal} {J. Phys. A}\ }\textbf {\bibinfo {volume} {24}},\ \bibinfo {pages} {4027} (\bibinfo {year} {1991})}\BibitemShut {NoStop}%
\bibitem [{\citenamefont {Cromwell}\ \emph {et~al.}(1998)\citenamefont {Cromwell}, \citenamefont {Beltrami},\ and\ \citenamefont {Rampichini}}]{cromwell1998borromean}%
  \BibitemOpen
  \bibfield  {author} {\bibinfo {author} {\bibfnamefont {P.}~\bibnamefont {Cromwell}}, \bibinfo {author} {\bibfnamefont {E.}~\bibnamefont {Beltrami}},\ and\ \bibinfo {author} {\bibfnamefont {M.}~\bibnamefont {Rampichini}},\ }\bibfield  {title} {\bibinfo {title} {The borromean rings},\ }\href@noop {} {\bibfield  {journal} {\bibinfo  {journal} {Math. Intell.}\ }\textbf {\bibinfo {volume} {20}},\ \bibinfo {pages} {53} (\bibinfo {year} {1998})}\BibitemShut {NoStop}%
\bibitem [{\citenamefont {Kempton}(2016)}]{kemptonARX16}%
  \BibitemOpen
  \bibfield  {author} {\bibinfo {author} {\bibfnamefont {M.}~\bibnamefont {Kempton}},\ }\href {https://arxiv.org/abs/1610.04672} {\bibinfo {title} {A non-backtracking polya's theorem}} (\bibinfo {year} {2016}),\ \Eprint {https://arxiv.org/abs/1610.04672} {arXiv:1610.04672 [math.CO]} \BibitemShut {NoStop}%
\bibitem [{Note9()}]{Note9}%
  \BibitemOpen
  \bibinfo {note} {As a concrete example, a trajectory which encloses one hexagon clockwise and a neighboring one counterclockwise is associated with a Brunnian vortex, yet $\protect \mathcal {Q}^{(2)} = 0$.}\BibitemShut {Stop}%
\bibitem [{\citenamefont {Sonin}(1997)}]{soninPRB97}%
  \BibitemOpen
  \bibfield  {author} {\bibinfo {author} {\bibfnamefont {E.~B.}\ \bibnamefont {Sonin}},\ }\bibfield  {title} {\bibinfo {title} {Magnus force in superfluids and superconductors},\ }\href {https://link.aps.org/doi/10.1103/PhysRevB.55.485} {\bibfield  {journal} {\bibinfo  {journal} {Phys. Rev. B}\ }\textbf {\bibinfo {volume} {55}},\ \bibinfo {pages} {485} (\bibinfo {year} {1997})}\BibitemShut {NoStop}%
\bibitem [{\citenamefont {Kobayashi}\ \emph {et~al.}(2019)\citenamefont {Kobayashi}, \citenamefont {Eto},\ and\ \citenamefont {Nitta}}]{kobayashiPRL19}%
  \BibitemOpen
  \bibfield  {author} {\bibinfo {author} {\bibfnamefont {M.}~\bibnamefont {Kobayashi}}, \bibinfo {author} {\bibfnamefont {M.}~\bibnamefont {Eto}},\ and\ \bibinfo {author} {\bibfnamefont {M.}~\bibnamefont {Nitta}},\ }\bibfield  {title} {\bibinfo {title} {Berezinskii-kosterlitz-thouless transition of two-component bose mixtures with intercomponent josephson coupling},\ }\href {https://link.aps.org/doi/10.1103/PhysRevLett.123.075303} {\bibfield  {journal} {\bibinfo  {journal} {Phys. Rev. Lett.}\ }\textbf {\bibinfo {volume} {123}},\ \bibinfo {pages} {075303} (\bibinfo {year} {2019})}\BibitemShut {NoStop}%
\bibitem [{\citenamefont {Lesser}\ \emph {et~al.}(2026)\citenamefont {Lesser}, \citenamefont {Huang}, \citenamefont {Sethna},\ and\ \citenamefont {Kim}}]{lesserPRB26}%
  \BibitemOpen
  \bibfield  {author} {\bibinfo {author} {\bibfnamefont {O.}~\bibnamefont {Lesser}}, \bibinfo {author} {\bibfnamefont {C.}~\bibnamefont {Huang}}, \bibinfo {author} {\bibfnamefont {J.~P.}\ \bibnamefont {Sethna}},\ and\ \bibinfo {author} {\bibfnamefont {E.-A.}\ \bibnamefont {Kim}},\ }\bibfield  {title} {\bibinfo {title} {Emblems of pair density waves: Dual identity of topological defects and their transport signatures},\ }\href {https://link.aps.org/doi/10.1103/1gyj-r5fr} {\bibfield  {journal} {\bibinfo  {journal} {Phys. Rev. B}\ }\textbf {\bibinfo {volume} {113}},\ \bibinfo {pages} {214505} (\bibinfo {year} {2026})}\BibitemShut {NoStop}%
\bibitem [{\citenamefont {Jones}\ \emph {et~al.}(2020)\citenamefont {Jones}, \citenamefont {Zou}, \citenamefont {Zhang},\ and\ \citenamefont {Tserkovnyak}}]{jonesPRB20}%
  \BibitemOpen
  \bibfield  {author} {\bibinfo {author} {\bibfnamefont {D.}~\bibnamefont {Jones}}, \bibinfo {author} {\bibfnamefont {J.}~\bibnamefont {Zou}}, \bibinfo {author} {\bibfnamefont {S.}~\bibnamefont {Zhang}},\ and\ \bibinfo {author} {\bibfnamefont {Y.}~\bibnamefont {Tserkovnyak}},\ }\bibfield  {title} {\bibinfo {title} {Energy storage in magnetic textures driven by vorticity flow},\ }\href {https://link.aps.org/doi/10.1103/PhysRevB.102.140411} {\bibfield  {journal} {\bibinfo  {journal} {Phys. Rev. B}\ }\textbf {\bibinfo {volume} {102}},\ \bibinfo {pages} {140411 (R)} (\bibinfo {year} {2020})}\BibitemShut {NoStop}%
\bibitem [{\citenamefont {Sato}\ and\ \citenamefont {Packard}(2011)}]{satoRPP11}%
  \BibitemOpen
  \bibfield  {author} {\bibinfo {author} {\bibfnamefont {Y.}~\bibnamefont {Sato}}\ and\ \bibinfo {author} {\bibfnamefont {R.~E.}\ \bibnamefont {Packard}},\ }\bibfield  {title} {\bibinfo {title} {Superfluid helium quantum interference devices: physics and applications},\ }\href {https://doi.org/10.1088/0034-4885/75/1/016401} {\bibfield  {journal} {\bibinfo  {journal} {Rep. Prog. Phys.}\ }\textbf {\bibinfo {volume} {75}},\ \bibinfo {pages} {016401} (\bibinfo {year} {2011})}\BibitemShut {NoStop}%
\bibitem [{\citenamefont {Coleman}(2015)}]{coleman15}%
  \BibitemOpen
  \bibfield  {author} {\bibinfo {author} {\bibfnamefont {P.}~\bibnamefont {Coleman}},\ }\href@noop {} {\emph {\bibinfo {title} {Introduction to many-body physics}}}\ (\bibinfo  {publisher} {Cambridge University Press},\ \bibinfo {year} {2015})\BibitemShut {NoStop}%
\bibitem [{Note10()}]{Note10}%
  \BibitemOpen
  \bibinfo {note} {To couple to the relative winding $\protect \boldsymbol {\nabla }\theta _{12}$, we would require an interaction of the Lifshitz form, which breaks inversion symmetry and has been discussed in the context of multiband superconductors~\cite {nagashimaPRB25}.}\BibitemShut {Stop}%
\bibitem [{\citenamefont {Zhu}\ \emph {et~al.}(2025)\citenamefont {Zhu}, \citenamefont {Kleinherbers}, \citenamefont {Levitov},\ and\ \citenamefont {Tserkovnyak}}]{zhuPRB25}%
  \BibitemOpen
  \bibfield  {author} {\bibinfo {author} {\bibfnamefont {Y.}~\bibnamefont {Zhu}}, \bibinfo {author} {\bibfnamefont {E.}~\bibnamefont {Kleinherbers}}, \bibinfo {author} {\bibfnamefont {L.}~\bibnamefont {Levitov}},\ and\ \bibinfo {author} {\bibfnamefont {Y.}~\bibnamefont {Tserkovnyak}},\ }\bibfield  {title} {\bibinfo {title} {Proposal for spin superfluid quantum interference device},\ }\href {https://link.aps.org/doi/10.1103/j8rt-c6qp} {\bibfield  {journal} {\bibinfo  {journal} {Phys. Rev. B}\ }\textbf {\bibinfo {volume} {112}},\ \bibinfo {pages} {L100405} (\bibinfo {year} {2025})}\BibitemShut {NoStop}%
\bibitem [{\citenamefont {Taylor}\ \emph {et~al.}(1970)\citenamefont {Taylor}, \citenamefont {Fergason},\ and\ \citenamefont {Arora}}]{taylorPRL70}%
  \BibitemOpen
  \bibfield  {author} {\bibinfo {author} {\bibfnamefont {T.~R.}\ \bibnamefont {Taylor}}, \bibinfo {author} {\bibfnamefont {J.~L.}\ \bibnamefont {Fergason}},\ and\ \bibinfo {author} {\bibfnamefont {S.~L.}\ \bibnamefont {Arora}},\ }\bibfield  {title} {\bibinfo {title} {Biaxial liquid crystals},\ }\href {https://link.aps.org/doi/10.1103/PhysRevLett.24.359} {\bibfield  {journal} {\bibinfo  {journal} {Phys. Rev. Lett.}\ }\textbf {\bibinfo {volume} {24}},\ \bibinfo {pages} {359} (\bibinfo {year} {1970})}\BibitemShut {NoStop}%
\bibitem [{\citenamefont {Xiao}\ \emph {et~al.}(2022)\citenamefont {Xiao}, \citenamefont {Borgh}, \citenamefont {Blinova}, \citenamefont {Ollikainen}, \citenamefont {Ruostekoski},\ and\ \citenamefont {Hall}}]{xiaoNATC22}%
  \BibitemOpen
  \bibfield  {author} {\bibinfo {author} {\bibfnamefont {Y.}~\bibnamefont {Xiao}}, \bibinfo {author} {\bibfnamefont {M.~O.}\ \bibnamefont {Borgh}}, \bibinfo {author} {\bibfnamefont {A.}~\bibnamefont {Blinova}}, \bibinfo {author} {\bibfnamefont {T.}~\bibnamefont {Ollikainen}}, \bibinfo {author} {\bibfnamefont {J.}~\bibnamefont {Ruostekoski}},\ and\ \bibinfo {author} {\bibfnamefont {D.~S.}\ \bibnamefont {Hall}},\ }\bibfield  {title} {\bibinfo {title} {Topological superfluid defects with discrete point group symmetries},\ }\href {https://doi.org/10.1038/s41467-022-32362-5} {\bibfield  {journal} {\bibinfo  {journal} {Nat. Commun.}\ }\textbf {\bibinfo {volume} {13}},\ \bibinfo {pages} {4635} (\bibinfo {year} {2022})}\BibitemShut {NoStop}%
\bibitem [{\citenamefont {Borgh}\ and\ \citenamefont {Ruostekoski}(2016)}]{borghPRL16}%
  \BibitemOpen
  \bibfield  {author} {\bibinfo {author} {\bibfnamefont {M.~O.}\ \bibnamefont {Borgh}}\ and\ \bibinfo {author} {\bibfnamefont {J.}~\bibnamefont {Ruostekoski}},\ }\bibfield  {title} {\bibinfo {title} {Core structure and non-abelian reconnection of defects in a biaxial nematic spin-2 bose-einstein condensate},\ }\href {https://link.aps.org/doi/10.1103/PhysRevLett.117.275302} {\bibfield  {journal} {\bibinfo  {journal} {Phys. Rev. Lett.}\ }\textbf {\bibinfo {volume} {117}},\ \bibinfo {pages} {275302} (\bibinfo {year} {2016})}\BibitemShut {NoStop}%
\bibitem [{\citenamefont {Semenoff}\ and\ \citenamefont {Zhou}(2007)}]{semenoffPRL07}%
  \BibitemOpen
  \bibfield  {author} {\bibinfo {author} {\bibfnamefont {G.~W.}\ \bibnamefont {Semenoff}}\ and\ \bibinfo {author} {\bibfnamefont {F.}~\bibnamefont {Zhou}},\ }\bibfield  {title} {\bibinfo {title} {Discrete symmetries and $1/3$--quantum vortices in condensates of $f=2$ cold atoms},\ }\href {https://link.aps.org/doi/10.1103/PhysRevLett.98.100401} {\bibfield  {journal} {\bibinfo  {journal} {Phys. Rev. Lett.}\ }\textbf {\bibinfo {volume} {98}},\ \bibinfo {pages} {100401} (\bibinfo {year} {2007})}\BibitemShut {NoStop}%
\bibitem [{Note11()}]{Note11}%
  \BibitemOpen
  \bibinfo {note} {To be precise, the truly nonabelian structure is determined by the quotient of the fundamental group with its (abelian) center. In this work, we have that $\pi _1(\protect \mathcal {M})/Z[\pi _1(\protect \mathcal {M})] = P_3/Z(P_3) = F_2$, which is an infinite-sized group.}\BibitemShut {Stop}%
\bibitem [{\citenamefont {Rybakov}(2026)}]{rybakovPRI26}%
  \BibitemOpen
  \bibfield  {author} {\bibinfo {author} {\bibfnamefont {F.~N.}\ \bibnamefont {Rybakov}},\ }\href@noop {} {} (\bibinfo {year} {2026}),\ \bibinfo {note} {private communication}\BibitemShut {NoStop}%
\bibitem [{\citenamefont {Garaud}\ \emph {et~al.}(2013)\citenamefont {Garaud}, \citenamefont {Carlstr\"om}, \citenamefont {Babaev},\ and\ \citenamefont {Speight}}]{garaudPRB13}%
  \BibitemOpen
  \bibfield  {author} {\bibinfo {author} {\bibfnamefont {J.}~\bibnamefont {Garaud}}, \bibinfo {author} {\bibfnamefont {J.}~\bibnamefont {Carlstr\"om}}, \bibinfo {author} {\bibfnamefont {E.}~\bibnamefont {Babaev}},\ and\ \bibinfo {author} {\bibfnamefont {M.}~\bibnamefont {Speight}},\ }\bibfield  {title} {\bibinfo {title} {Chiral $\mathbb{C}{P}^{2}$ skyrmions in three-band superconductors},\ }\href {https://link.aps.org/doi/10.1103/PhysRevB.87.014507} {\bibfield  {journal} {\bibinfo  {journal} {Phys. Rev. B}\ }\textbf {\bibinfo {volume} {87}},\ \bibinfo {pages} {014507} (\bibinfo {year} {2013})}\BibitemShut {NoStop}%
\bibitem [{\citenamefont {Zheng}\ \emph {et~al.}(2026)\citenamefont {Zheng}, \citenamefont {Hu}, \citenamefont {Yu}, \citenamefont {Ji}, \citenamefont {Timoshuk}, \citenamefont {Garaud}, \citenamefont {Xu}, \citenamefont {Li}, \citenamefont {Gao}, \citenamefont {Lu}, \citenamefont {Grinenko}, \citenamefont {Babaev}, \citenamefont {Yuan}, \citenamefont {Wu}, \citenamefont {Lv}, \citenamefont {Yim},\ and\ \citenamefont {Ding}}]{zhengSCI26}%
  \BibitemOpen
  \bibfield  {author} {\bibinfo {author} {\bibfnamefont {Y.}~\bibnamefont {Zheng}}, \bibinfo {author} {\bibfnamefont {Q.}~\bibnamefont {Hu}}, \bibinfo {author} {\bibfnamefont {X.}~\bibnamefont {Yu}}, \bibinfo {author} {\bibfnamefont {H.}~\bibnamefont {Ji}}, \bibinfo {author} {\bibfnamefont {I.}~\bibnamefont {Timoshuk}}, \bibinfo {author} {\bibfnamefont {J.}~\bibnamefont {Garaud}}, \bibinfo {author} {\bibfnamefont {H.}~\bibnamefont {Xu}}, \bibinfo {author} {\bibfnamefont {Y.}~\bibnamefont {Li}}, \bibinfo {author} {\bibfnamefont {Y.}~\bibnamefont {Gao}}, \bibinfo {author} {\bibfnamefont {X.}~\bibnamefont {Lu}}, \bibinfo {author} {\bibfnamefont {V.}~\bibnamefont {Grinenko}}, \bibinfo {author} {\bibfnamefont {E.}~\bibnamefont {Babaev}}, \bibinfo {author} {\bibfnamefont {N.~F.~Q.}\ \bibnamefont {Yuan}}, \bibinfo {author} {\bibfnamefont {R.}~\bibnamefont {Wu}}, \bibinfo {author} {\bibfnamefont {B.}~\bibnamefont {Lv}}, \bibinfo {author} {\bibfnamefont {C.-M.}\ \bibnamefont {Yim}},\ and\ \bibinfo {author}
  {\bibfnamefont {H.}~\bibnamefont {Ding}},\ }\bibfield  {title} {\bibinfo {title} {Observation of quantum vortex core fractionalization and skyrmion formation in a superconductor},\ }\href {https://www.science.org/doi/abs/10.1126/science.ads0189} {\bibfield  {journal} {\bibinfo  {journal} {Science}\ }\textbf {\bibinfo {volume} {393}},\ \bibinfo {pages} {80} (\bibinfo {year} {2026})}\BibitemShut {NoStop}%
\bibitem [{\citenamefont {Kim}\ \emph {et~al.}(2022)\citenamefont {Kim}, \citenamefont {Beach}, \citenamefont {Lee}, \citenamefont {Ono}, \citenamefont {Rasing},\ and\ \citenamefont {Yang}}]{kimNATM22}%
  \BibitemOpen
  \bibfield  {author} {\bibinfo {author} {\bibfnamefont {S.~K.}\ \bibnamefont {Kim}}, \bibinfo {author} {\bibfnamefont {G.~S.~D.}\ \bibnamefont {Beach}}, \bibinfo {author} {\bibfnamefont {K.-J.}\ \bibnamefont {Lee}}, \bibinfo {author} {\bibfnamefont {T.}~\bibnamefont {Ono}}, \bibinfo {author} {\bibfnamefont {T.}~\bibnamefont {Rasing}},\ and\ \bibinfo {author} {\bibfnamefont {H.}~\bibnamefont {Yang}},\ }\bibfield  {title} {\bibinfo {title} {Ferrimagnetic spintronics},\ }\href {https://doi.org/10.1038/s41563-021-01139-4} {\bibfield  {journal} {\bibinfo  {journal} {Nat. Mater.}\ }\textbf {\bibinfo {volume} {21}},\ \bibinfo {pages} {24} (\bibinfo {year} {2022})}\BibitemShut {NoStop}%
\bibitem [{\citenamefont {Quessab}\ \emph {et~al.}(2020)\citenamefont {Quessab}, \citenamefont {Xu}, \citenamefont {Ma}, \citenamefont {Zhou}, \citenamefont {Riley}, \citenamefont {Shaw}, \citenamefont {Nembach}, \citenamefont {Poon},\ and\ \citenamefont {Kent}}]{quessabSCIR20}%
  \BibitemOpen
  \bibfield  {author} {\bibinfo {author} {\bibfnamefont {Y.}~\bibnamefont {Quessab}}, \bibinfo {author} {\bibfnamefont {J.-W.}\ \bibnamefont {Xu}}, \bibinfo {author} {\bibfnamefont {C.~T.}\ \bibnamefont {Ma}}, \bibinfo {author} {\bibfnamefont {W.}~\bibnamefont {Zhou}}, \bibinfo {author} {\bibfnamefont {G.~A.}\ \bibnamefont {Riley}}, \bibinfo {author} {\bibfnamefont {J.~M.}\ \bibnamefont {Shaw}}, \bibinfo {author} {\bibfnamefont {H.~T.}\ \bibnamefont {Nembach}}, \bibinfo {author} {\bibfnamefont {S.~J.}\ \bibnamefont {Poon}},\ and\ \bibinfo {author} {\bibfnamefont {A.~D.}\ \bibnamefont {Kent}},\ }\bibfield  {title} {\bibinfo {title} {Tuning interfacial dzyaloshinskii-moriya interactions in thin amorphous ferrimagnetic alloys},\ }\href {https://doi.org/10.1038/s41598-020-64427-0} {\bibfield  {journal} {\bibinfo  {journal} {Sci. Rep.}\ }\textbf {\bibinfo {volume} {10}},\ \bibinfo {pages} {7447} (\bibinfo {year} {2020})}\BibitemShut {NoStop}%
\bibitem [{\citenamefont {Zou}\ \emph {et~al.}(2023)\citenamefont {Zou}, \citenamefont {Bosco}, \citenamefont {Pal}, \citenamefont {Parkin}, \citenamefont {Klinovaja},\ and\ \citenamefont {Loss}}]{zouPRR23}%
  \BibitemOpen
  \bibfield  {author} {\bibinfo {author} {\bibfnamefont {J.}~\bibnamefont {Zou}}, \bibinfo {author} {\bibfnamefont {S.}~\bibnamefont {Bosco}}, \bibinfo {author} {\bibfnamefont {B.}~\bibnamefont {Pal}}, \bibinfo {author} {\bibfnamefont {S.~S.~P.}\ \bibnamefont {Parkin}}, \bibinfo {author} {\bibfnamefont {J.}~\bibnamefont {Klinovaja}},\ and\ \bibinfo {author} {\bibfnamefont {D.}~\bibnamefont {Loss}},\ }\bibfield  {title} {\bibinfo {title} {Quantum computing on magnetic racetracks with flying domain wall qubits},\ }\href {https://link.aps.org/doi/10.1103/PhysRevResearch.5.033166} {\bibfield  {journal} {\bibinfo  {journal} {Phys. Rev. Res.}\ }\textbf {\bibinfo {volume} {5}},\ \bibinfo {pages} {033166} (\bibinfo {year} {2023})}\BibitemShut {NoStop}%
\bibitem [{\citenamefont {Wan}\ and\ \citenamefont {Tchernyshyov}(2012)}]{wanPRL12}%
  \BibitemOpen
  \bibfield  {author} {\bibinfo {author} {\bibfnamefont {Y.}~\bibnamefont {Wan}}\ and\ \bibinfo {author} {\bibfnamefont {O.}~\bibnamefont {Tchernyshyov}},\ }\bibfield  {title} {\bibinfo {title} {Quantum strings in quantum spin ice},\ }\href {https://link.aps.org/doi/10.1103/PhysRevLett.108.247210} {\bibfield  {journal} {\bibinfo  {journal} {Phys. Rev. Lett.}\ }\textbf {\bibinfo {volume} {108}},\ \bibinfo {pages} {247210} (\bibinfo {year} {2012})}\BibitemShut {NoStop}%
\bibitem [{\citenamefont {Pan}\ \emph {et~al.}(2016)\citenamefont {Pan}, \citenamefont {Laurita}, \citenamefont {Ross}, \citenamefont {Gaulin},\ and\ \citenamefont {Armitage}}]{panNATP16}%
  \BibitemOpen
  \bibfield  {author} {\bibinfo {author} {\bibfnamefont {L.}~\bibnamefont {Pan}}, \bibinfo {author} {\bibfnamefont {N.~J.}\ \bibnamefont {Laurita}}, \bibinfo {author} {\bibfnamefont {K.~A.}\ \bibnamefont {Ross}}, \bibinfo {author} {\bibfnamefont {B.~D.}\ \bibnamefont {Gaulin}},\ and\ \bibinfo {author} {\bibfnamefont {N.~P.}\ \bibnamefont {Armitage}},\ }\bibfield  {title} {\bibinfo {title} {A measure of monopole inertia in the quantum spin ice yb2ti2o7},\ }\href {https://doi.org/10.1038/nphys3608} {\bibfield  {journal} {\bibinfo  {journal} {Nat. Phys.}\ }\textbf {\bibinfo {volume} {12}},\ \bibinfo {pages} {361} (\bibinfo {year} {2016})}\BibitemShut {NoStop}%
\bibitem [{\citenamefont {Lee}(2010)}]{leeJCM10}%
  \BibitemOpen
  \bibfield  {author} {\bibinfo {author} {\bibfnamefont {E.-K.}\ \bibnamefont {Lee}},\ }\bibfield  {title} {\bibinfo {title} {A positive presentation for the pure braid group},\ }\href {https://koreascience.or.kr/article/JAKO201007648745187.page} {\bibfield  {journal} {\bibinfo  {journal} {J. Chungcheong Math. Soc.}\ }\textbf {\bibinfo {volume} {23}},\ \bibinfo {pages} {555} (\bibinfo {year} {2010})}\BibitemShut {NoStop}%
\bibitem [{\citenamefont {Vladimir}(1969)}]{vladimir1969arnol}%
  \BibitemOpen
  \bibfield  {author} {\bibinfo {author} {\bibfnamefont {I.~A.}\ \bibnamefont {Vladimir}},\ }\bibfield  {title} {\bibinfo {title} {The cohomology ring of the group of dyed braids},\ }\href@noop {} {\bibfield  {journal} {\bibinfo  {journal} {Mat. Zametki}\ }\textbf {\bibinfo {volume} {5}},\ \bibinfo {pages} {227} (\bibinfo {year} {1969})}\BibitemShut {NoStop}%
\bibitem [{\citenamefont {Ramsay}(2025)}]{ramsayJCA25}%
  \BibitemOpen
  \bibfield  {author} {\bibinfo {author} {\bibfnamefont {C.}~\bibnamefont {Ramsay}},\ }\bibfield  {title} {\bibinfo {title} {Listing words in free groups},\ }\href {https://www.sciencedirect.com/science/article/pii/S2772827725000051} {\bibfield  {journal} {\bibinfo  {journal} {J. Comput. Algebra}\ }\textbf {\bibinfo {volume} {13-14}},\ \bibinfo {pages} {100034} (\bibinfo {year} {2025})}\BibitemShut {NoStop}%
\bibitem [{\citenamefont {Nagashima}\ \emph {et~al.}(2025)\citenamefont {Nagashima}, \citenamefont {Mouilleron},\ and\ \citenamefont {Tsuji}}]{nagashimaPRB25}%
  \BibitemOpen
  \bibfield  {author} {\bibinfo {author} {\bibfnamefont {R.}~\bibnamefont {Nagashima}}, \bibinfo {author} {\bibfnamefont {T.}~\bibnamefont {Mouilleron}},\ and\ \bibinfo {author} {\bibfnamefont {N.}~\bibnamefont {Tsuji}},\ }\bibfield  {title} {\bibinfo {title} {Optically active higgs and leggett modes in multiband pair-density-wave superconductors with lifshitz invariant},\ }\href {https://link.aps.org/doi/10.1103/cgyd-2g11} {\bibfield  {journal} {\bibinfo  {journal} {Phys. Rev. B}\ }\textbf {\bibinfo {volume} {112}},\ \bibinfo {pages} {024503} (\bibinfo {year} {2025})}\BibitemShut {NoStop}%
\end{thebibliography}%

\end{document}